\documentclass[]{spie}
 
\usepackage{amsmath,amsfonts,amssymb}
\usepackage{graphicx}
\usepackage{subfigure}
\usepackage[colorlinks=true, allcolors=blue]{hyperref}
\usepackage{wrapfig}
\usepackage{caption}

\title{The Nancy Grace Roman Space Telescope Coronagraph Instrument (CGI) Technology Demonstration}
\author[a]{N. Jeremy Kasdin}
\author[b]{Vanessa P. Bailey}
\author[b]{Bertrand Mennesson}
\author[b]{Robert T. Zellem}
\author[b]{Marie Ygouf}
\author[b]{Jason Rhodes}
\author[b]{Thomas Luchik}
\author[b]{Feng Zhao}
\author[b]{A J Eldorado Riggs}
\author[b]{Byoung-Joon Seo}
\author[b]{John Krist}
\author[b]{Brian Kern}
\author[b]{Hong Tang}
\author[c]{Bijan Nemati}
\author[d]{Tyler D. Groff}
\author[d]{Neil Zimmerman}
\author[e]{Bruce Macintosh}
\author[f]{Margaret Turnbull}
\author[g]{John Debes}
\author[h]{Ewan S. Douglas}
\author[i]{Roxana E. Lupu}

\affil[a]{University of San Francisco,College of Arts and Sciences, 2130 Fulton St., San Francisco, CA, USA, 94117}
\affil[b]{Jet Propulsion Laboratory - California Institute of Technology, 4800 Oak Grove Dr, Pasadena, CA, USA, 91109}
\affil[c]{University of Alabama in Huntsville, 301 Sparkman Drive, Huntsville, AL, USA, 35899 }
\affil[d]{NASA Goddard Space Flight Center, 8800 Greenbelt Rd, Greenbelt, MD, USA, 20771}
\affil[e]{Stanford University, Stanford, CA, USA, 94305}
\affil[f]{The SETI Institute, 189 Bernardo Ave, Suite 200, Mountain View, CA, USA, 94043}
\affil[g]{Space Telescope Science Institute, 3700 San Martin Dr., Baltimore, MD,USA, 21218}
\affil[h]{Steward Observatory, 933 North Cherry Avenue, Tucson, AZ, USA, 85721-0065}
\affil[i]{BAER Institute/NASA Ames Research Center, Moffett Field, CA, USA, 94035}

\authorinfo{Further author information send correspondence to Jeremy Kasdin, E-mail: jkasdin@usfca.edu}

\begin{document}
\maketitle

\begin{abstract}
    The Coronagraph Instrument (CGI) on the Nancy Grace Roman Space Telescope will demonstrate the high-contrast technology necessary for visible-light exoplanet imaging and spectroscopy from space via direct imaging of Jupiter-size planets and debris disks.  This in-space experience is a critical step toward future, larger missions targeted at direct imaging of Earth-like planets in the habitable zones of nearby stars.  This paper presents an overview of the current instrument design and requirements, highlighting the critical hardware, algorithms, and operations being demonstrated. We also describe several exoplanet and circumstellar disk science cases enabled by these capabilities. A competitively selected Community Participation Program team will be  an integral part of the technology demonstration and could perform additional CGI observations beyond the initial tech demo if the instrument performance warrants it.
\end{abstract}


\section{Introduction}
This paper gives a brief overview of the Coronagraph Instrument on the Nancy Grace Roman Space Telescope.  The Roman Space Telescope (formerly called WFIRST) is NASA’s next flagship observatory, planned for launch in 2025, employing a 2.4 meter diameter telescope and two instruments: a Wide-Field Instrument (WFI) for cosmology and dark energy and the Coronagraph Instrument (CGI) technology demonstrator for performing high-contrast imaging.  This paper introduces the goals, design, and progress of CGI.  There are several more detailed talks on various aspects of the instrument in this and other sessions of the conference that will be referenced throughout.

\section{CGI Technology Demonstration}
\begin{figure}[h]
   \begin{center}
   \includegraphics[height=8cm]{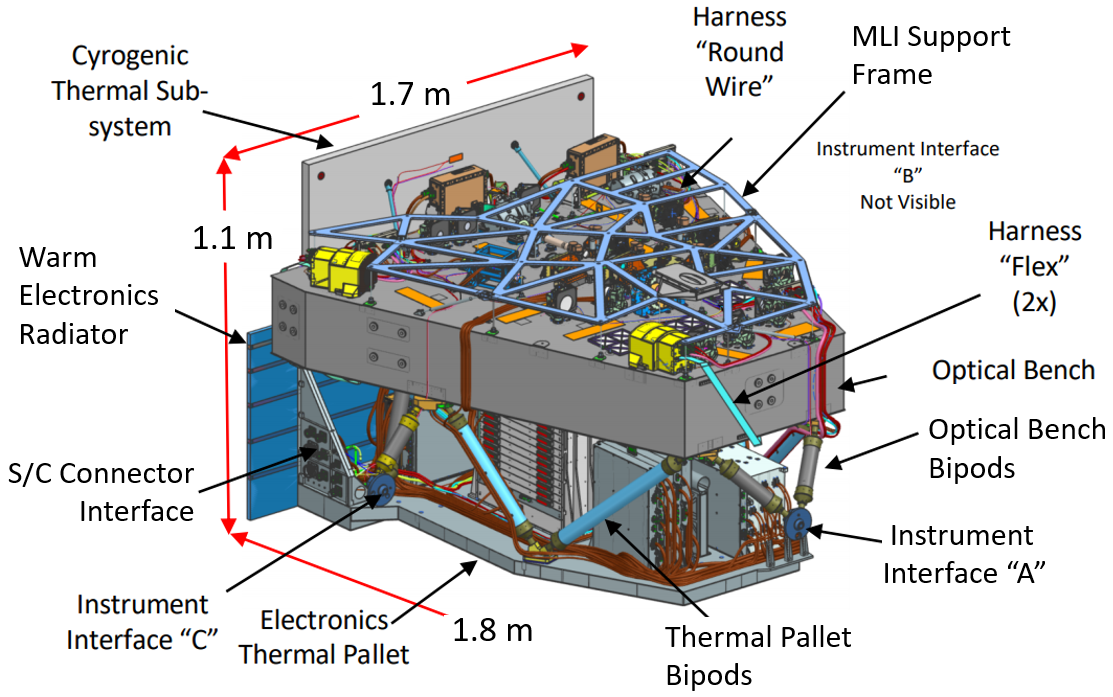}
   \end{center}
   \caption {\label{fig:cad} CAD diagram of the CGI instrument entering CDR.}
\end{figure}

The CGI instrument, whose schematic is shown in Figure~\ref{fig:cad}, is scheduled for its Critical Design Review (CDR) in April 2020.  More details on the overall optical layout are in the papers by Poberezhskiy \cite{Poberezhskiy20} and Tang\cite{Tang19}, and many specifications, such as filter and detector parameters, are publicly available.\footnote{\url{https://roman.ipac.caltech.edu/sims/Param_db.html}}  Currently, CGI is on schedule to meet the 2025 launch of Roman.  The primary purpose of CGI is to demonstrate the essential technologies necessary for high-contrast imaging that will be needed in future missions targeted at finding and characterizing Earth-like planets in the habitable zone of nearby stars\cite{Mennesson20} (such as the HabEx\cite{HabEx} and LUVOIR\cite{LUVOIR} mission studies).

As such, CGI has five primary technology demonstration objectives.  CGI will be the first coronagraphic instrument in space with active wavefront control, an important milestone in the evolution of high-contrast imaging and an essential precursor technology for all future large telescopes.  It will demonstrate and make flight-ready key hardware elements associated with advanced coronagraphy and active wavefront control as well as the algorithms for sensing, control, and data processing and analysis.  One of the most important things that will be learned from flying CGI is how the coronagraph interacts with the observatory as a whole and with the spacecraft control systems in particular, thus dramatically reducing the risk of future missions.  Finally, CGI will pioneer new advanced data processing techniques for extracting planets from noisy, photon-counting images, a key advancement for future missions.  

\section{Wavefront Sensing and Control for High Contrast}

\begin{figure}
   \begin{center}
   \includegraphics[height=8.5cm]{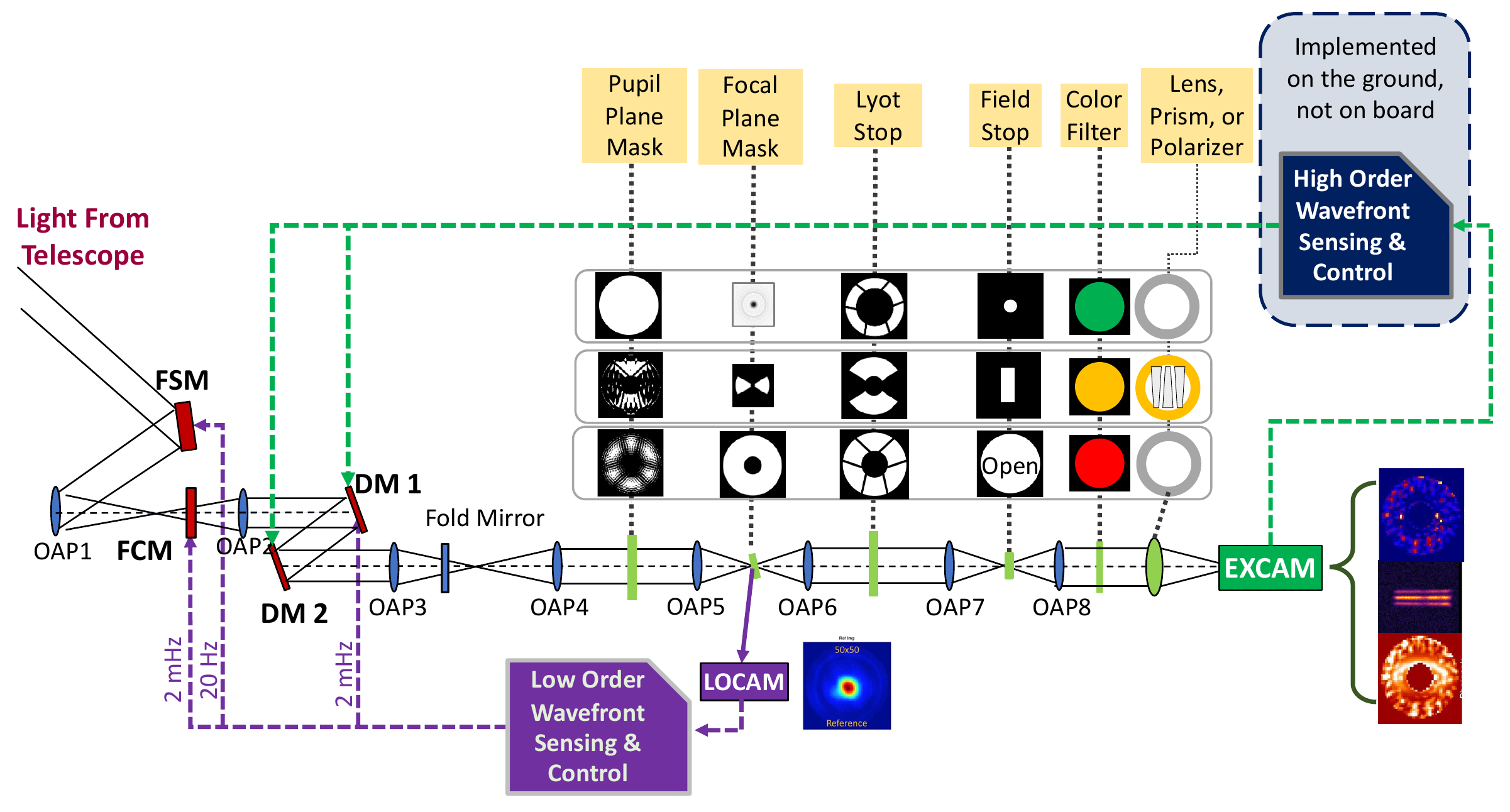}
   \end{center}
   \caption {\label{fig:WFSC} The CGI optical train and wavefront sensing and control architecture. The LOWFS/C loop operates with control bandwidths of $\sim$20~Hz and $\sim$2~mHz, for tip/tilt and Z4-Z11, respectively. HOWFS/C is not conducted continuously, but rather on the order of once per day, with processing offloaded to the ground.}
\end{figure}

The heart of the CGI technology demonstration is the active wavefront sensing and control system.  Figure~\ref{fig:WFSC} shows the architecture of the CGI sensing and control.  This figure illustrates how all the elements must work together as a system.  Low-order optical errors and fast variations (such as tip/tilt due to spacecraft pointing) are controlled by the Low-Order Wavefront Sensing and Control (LOWFS/C) loop, where aberrations and motion are measured using a Zernike wavefront sensor that uses starlight rejected by the coronagraph masks \cite{Shi2018, Shi16, Sidick19}.  The LOWFS/C loop operates with control bandwidths of $\sim$20~Hz and $\sim$2~mHz, for tip/tilt and Z4-Z11, respectively.   CGI's high-contrast performance is only guaranteed for stars of Vmag$\leq$5, as LOWFS SNR begins to degrade below this cutoff, particularly for the fast tip/tilt loop. Note that the feed-forward portion of the LOWFS/C tip/tilt loop described in Shi\cite{Shi2018} will not be implemented in flight. Expected tip/tilt performance is better than 1~mas RMS. The High-Order Wavefront Sensing and Control (HOWFS/C) loop uses multiple measurements at the final imaging camera with probe signals from the two deformable mirrors (DMs) for estimation\cite{Zhou19, Zhou20}.  While the nominal control approach uses the Electric Field Conjugation algorithm\cite{give2007electric} to determine the DM settings, the software is being configured to allow testing of various types of control approaches.  After CGI's preliminary design review (PDR), the decision was made to perform the slow, high-order control loop via telemetry and commands from the ground (Ground In The Loop, or GITL).  

Also shown in Figure~\ref{fig:WFSC} and described in Table~\ref{table:obsmodes} are the three primary modes of CGI: a short-wavelength mode with 10\% bandwidth\footnote{Bandwidth is the region where average transmission $>90\%$} centered at 575~nm (Band 1) for imaging and polarimetry from 3-9 $\lambda$/D using the Hybrid Lyot Coronagraph (HLC); a medium-wavelength, 15\% band centered at 730 nm (Band 3) for spectroscopy in a ``bowtie'' shaped field of view from 3-9~$\lambda$/D using the Shaped Pupil Coronagraph (SPC); and a 10\%, wide-field, long-wavelength band centered at 825~nm (Band 4) for imaging and polarimetry from 6--20~$\lambda$/D using a second Shaped Pupil Coronagraph. A fourth 15\% spectroscopy band (Band 2) centered at 660~nm is installed on CGI, but will not be tested on the ground, due to schedule constraints.

\begin{table}
\begin{center}
\begin{tabular}{ c c c c c c c}
 & $\lambda_0$ & FWHM & IWA & OWA & Angular  & Polarimetry \\
Name & [nm] & [nm] & [$\lambda/D$] & [$\lambda/D$] & Coverage [$^\circ$] & Compatible? \\
\hline
Band 1 imaging      & 575 & 70  & 3 & 9 & 360 & yes \\
Band 2 spectroscopy* & 660 & 112 & 3 & 9 & 2$\times$65 & no \\
Band 3 spectroscopy & 730 & 122 & 3 & 9 & 2$\times$65 & no \\
Band 4 imaging      & 825 & 94  & 6 & 20 & 360 & yes \\   
\end{tabular}
\caption{CGI observing modes. Note that Band 2 spectroscopy hardware will be installed, but will not be tested on the ground; hence, it is not an officially supported observing mode.}
\label{table:obsmodes}
\end{center}
\end{table}

\begin{figure}
  \begin{center}
   \begin{tabular}{ccc}
   \includegraphics[scale=0.8]{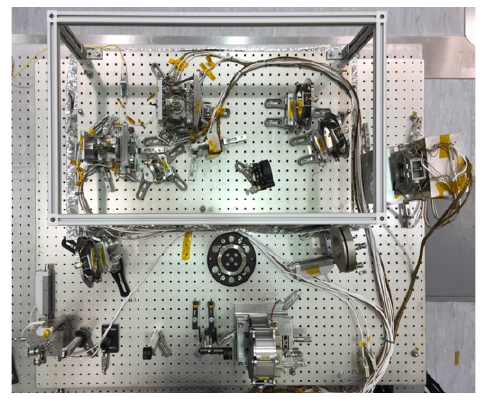} &
   \includegraphics[scale=0.9]{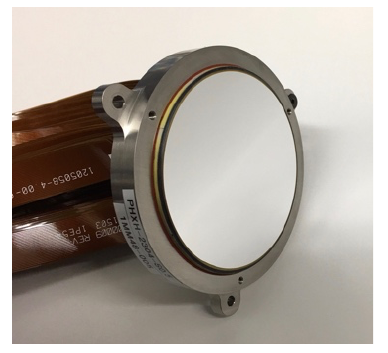} &
    \includegraphics[scale=0.4]{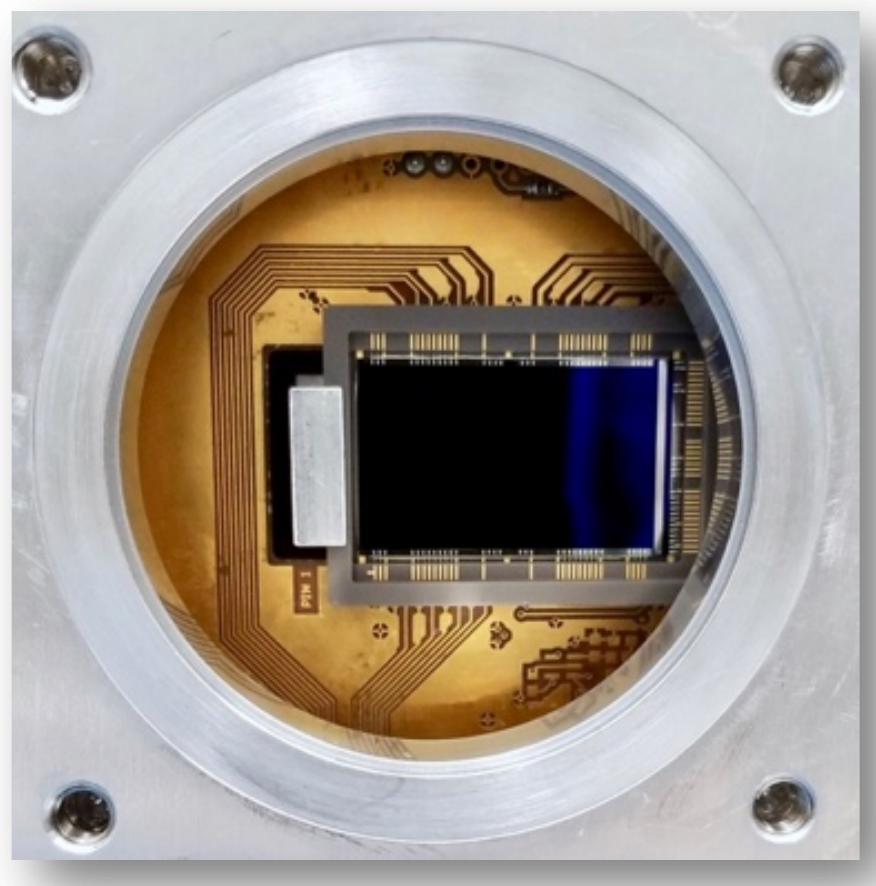} \\
    \includegraphics[scale=0.45]{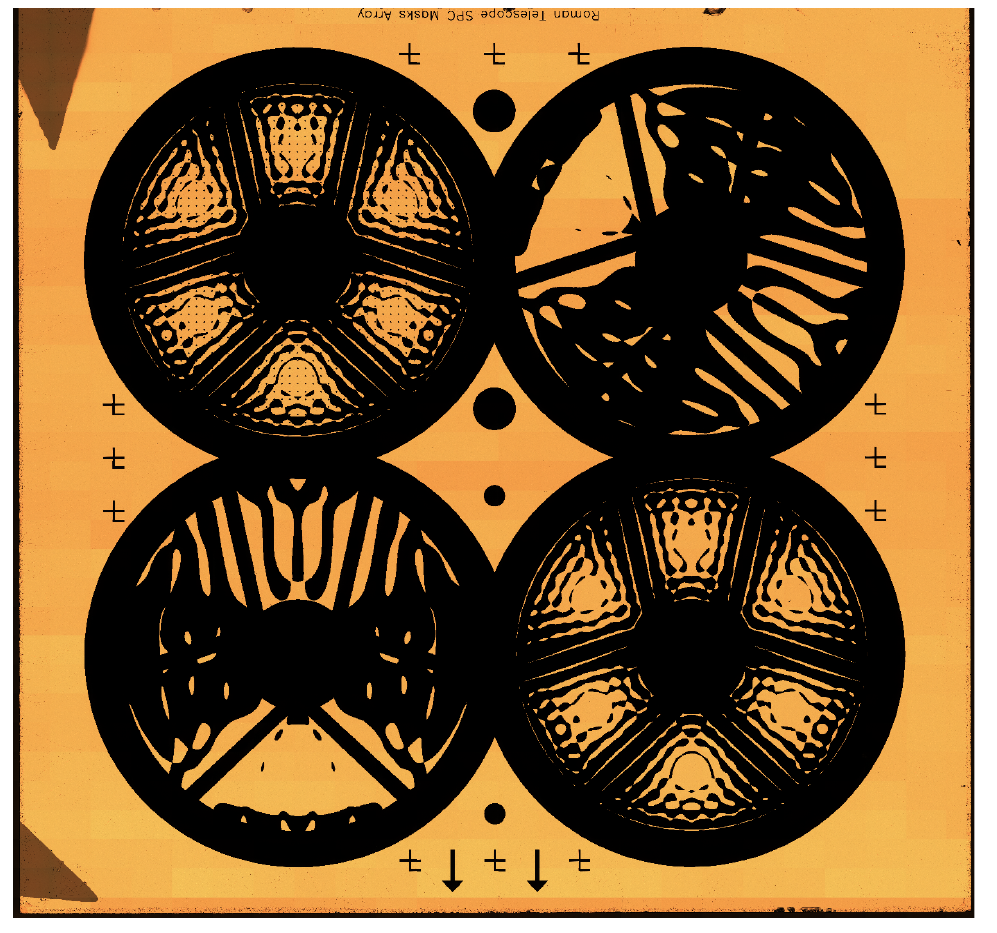} &
     \includegraphics[scale=0.35]{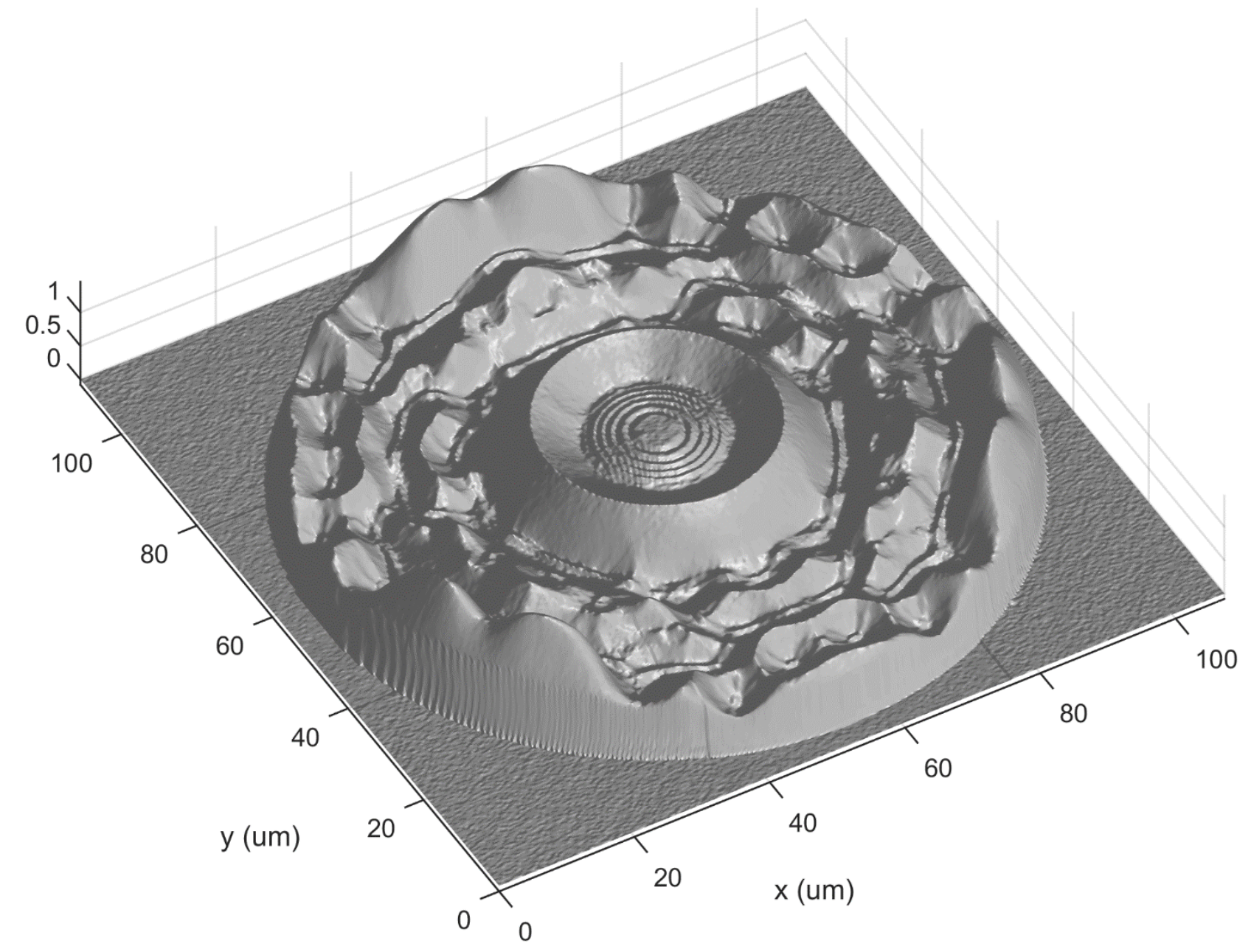} 
   \end{tabular}
   \end{center}
   \caption {\label{fig:hardware} New coronagraph technology elements being demonstrated on CGI.  Clockwise from upper left: Wavefront sensing and control optical layout, AOX Corp. 48x48 actuator Electrostrictive PMN (lead magnesium niobate) DM, Teledyne 1k x 1k EMCCD, Hybrid Lyot image plane mask, Black Silicon reflctive shaped pupils. }
\end{figure}

Successfully implementing the WFSC system above relies on several technology advances, all considered enabling for future high-contrast imaging missions\cite{Mennesson20}.  The key ones are shown in Figure~\ref{fig:hardware}. These include the algorithms and optical configuration needed for the wavefront sensing and control, which has been shown in the lab to be able to correct phase errors to better than 1 nm\cite{Poberezhskiy20, Zhou20}; new, large format deformable mirrors produced by AOA Xinetics, which have reached TRL 6 in the laboratory; masks and stops associated with the two types of coronagraphs being demonstrated on CGI, the Shaped Pupil Lyot Coronagraph (SPC\cite{Zimmerman16, Riggs17}) and the Hybrid Lyot Coronagraph (HLC\cite{Seo16, Trauger16}) (all flight designs are completed and the flight masks are in production\cite{Bala19}); extremely low-noise, electron-multiplying CCD detectors operated in photon-counting mode, necessary because of the very high magnitude of companion planets\cite{Bush20, Effinger18, Harding16};  and the Zernike wavefront sensor used in the low-order control loop, which has also been successfully demonstrated in the lab.\cite{Poberezhskiy20, Shi16, Shi2018}  All of these elements have passed all design milestones and flight article production has begun.

\begin{figure}
  \begin{center}
   \subfigure[Hybrid Lyot ]{\includegraphics[width=2.2 in]{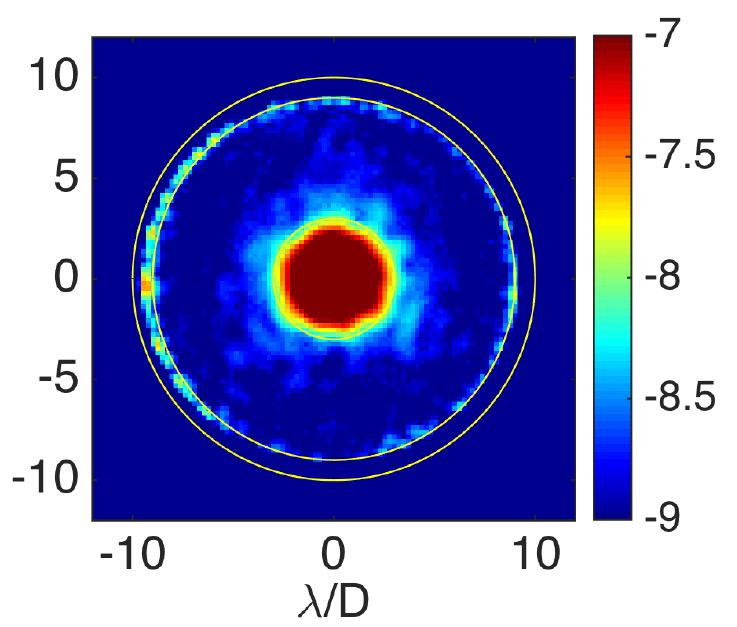}\label{fig:hlc_result}}
   \subfigure[``Bowtie'' Shaped Pupil Coronaraph]{\includegraphics[width=2.2 in]{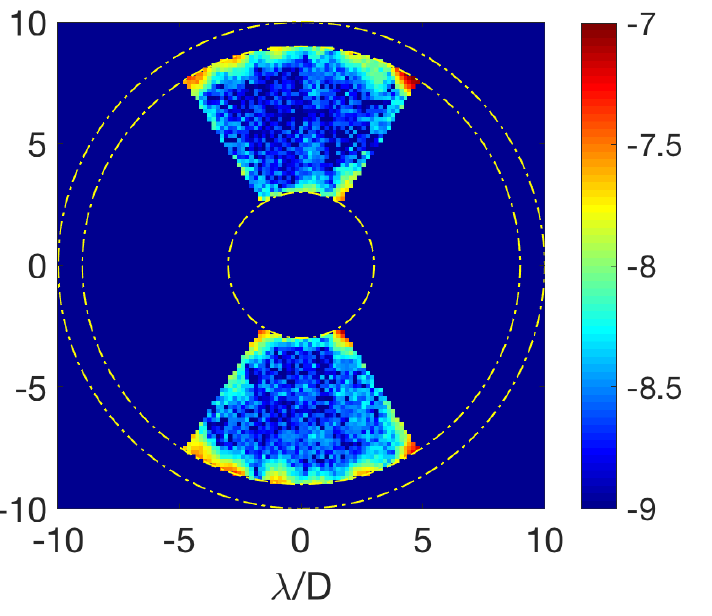}\label{fig:sp_result}}
  \subfigure[``Wide'' Shaped Pupil ]{\includegraphics[width=2.2 in]{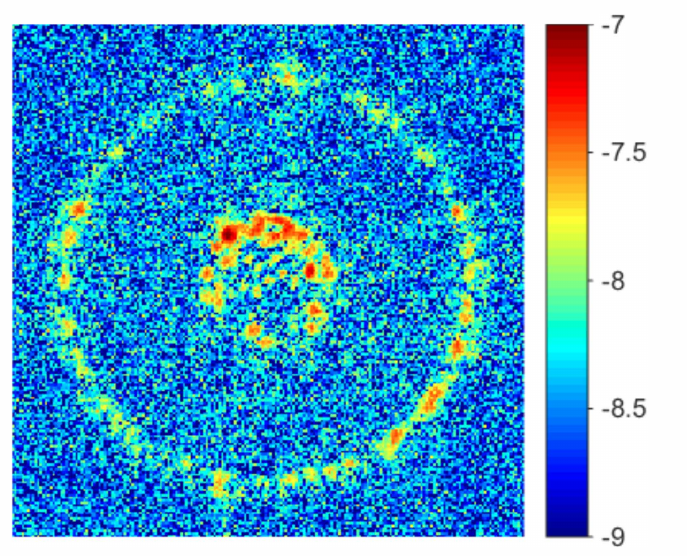}\label{wide_result}}
   \end{center}
   \caption {\label{fig:labresults} Demonstrations of high contrast in the three coronagraph modes in the JPL High Contrast Imaging Testbed with high-order and low-order wavefront control. (a) The Band 1 Hybrid Lyot Coronagraph in 10\% band around 550 nm in a 3-9 $\lambda$/D annulus with a mean contrast of $1.6\times10^{-9}$. (b) The Band 3 ``Bowtie'' Shaped Pupil Coronagraph in a 15\% band around 730 nm from 3-9 $\lambda$/D in two $65^\circ$ sections, with a mean contrast of $4.8\times10^{-9}$ .  (c) The Band 4 ``Wide field'' Shaped Pupil Coronagraph in a 10\% band around 825 nm from 6.5-20 $\lambda$/D with a mean contrast of $4.3\times10^{-9}$. }
\end{figure}

All three modes for CGI have been tested with flight-like masks, deformable mirrors and both high-order and low-order control in the High-Contrast Imaging Testbed at JPL\cite{Seo17, Seo18, Marx18, Cady17}.  As Figure~\ref{fig:labresults} shows, all three have achieved contrasts below $10^{-8}$ and reaching close to $10^{-9}$.  These tests are significant milestones toward CDR in April 2020 and give confidence in the performance predictions of CGI\cite{Zhou19, Nemati17}.  More detail can be found in the paper by Zhou.\cite{Zhou20}

\section{Spectroscopy and Polarization}

\begin{figure}
    \centering
    \subfigure[Alignment of slit with the planet.]{\includegraphics[scale=0.6]{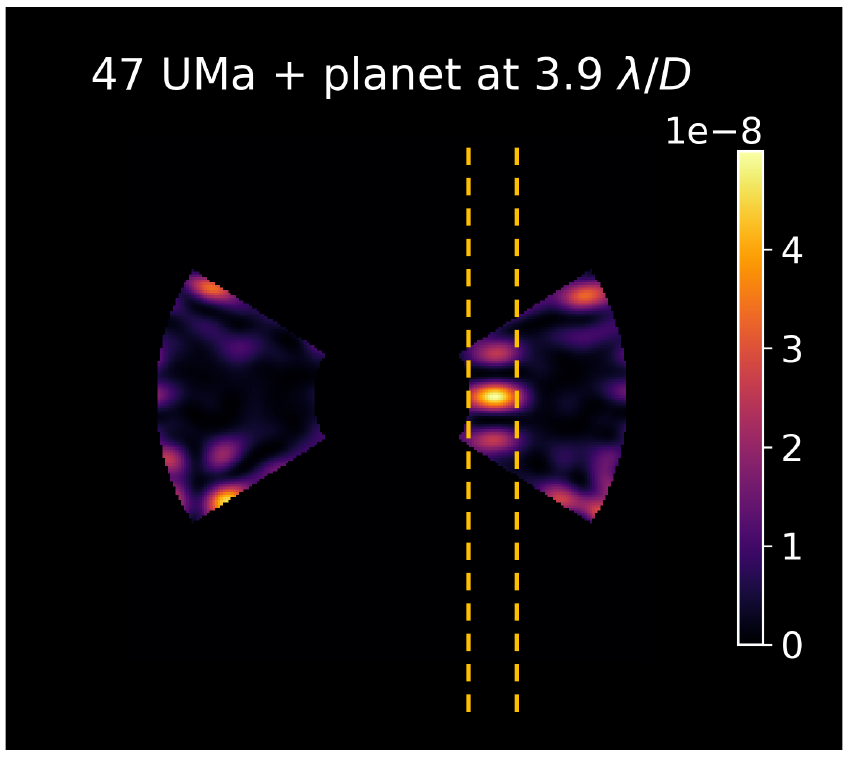}\label{fig:slitsim}}
    \subfigure[Dispersed image after post-processing]{\includegraphics[scale=0.6]{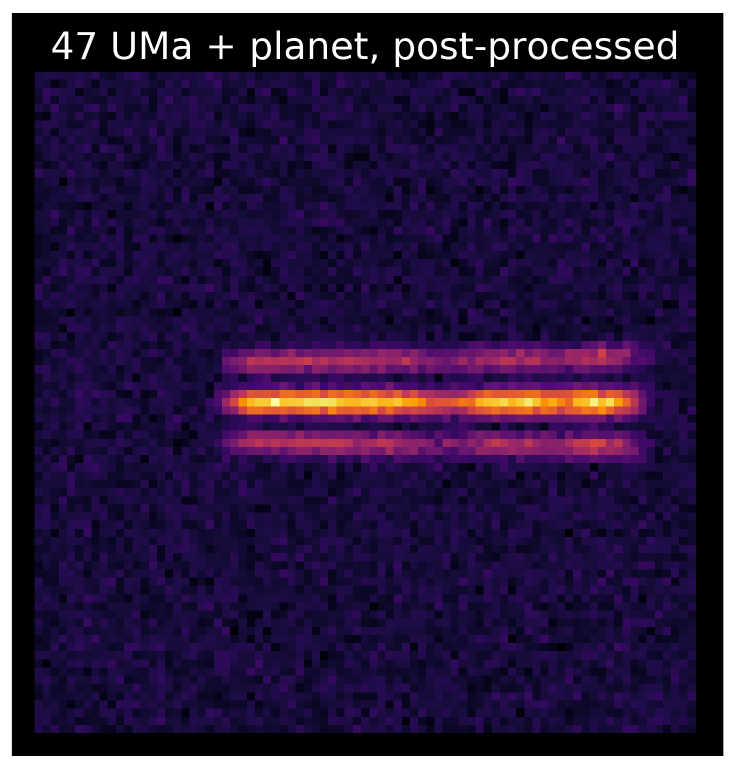}\label{fig:dispersion}}
    \subfigure[Extracted spectrum.]{\includegraphics[scale=0.6]{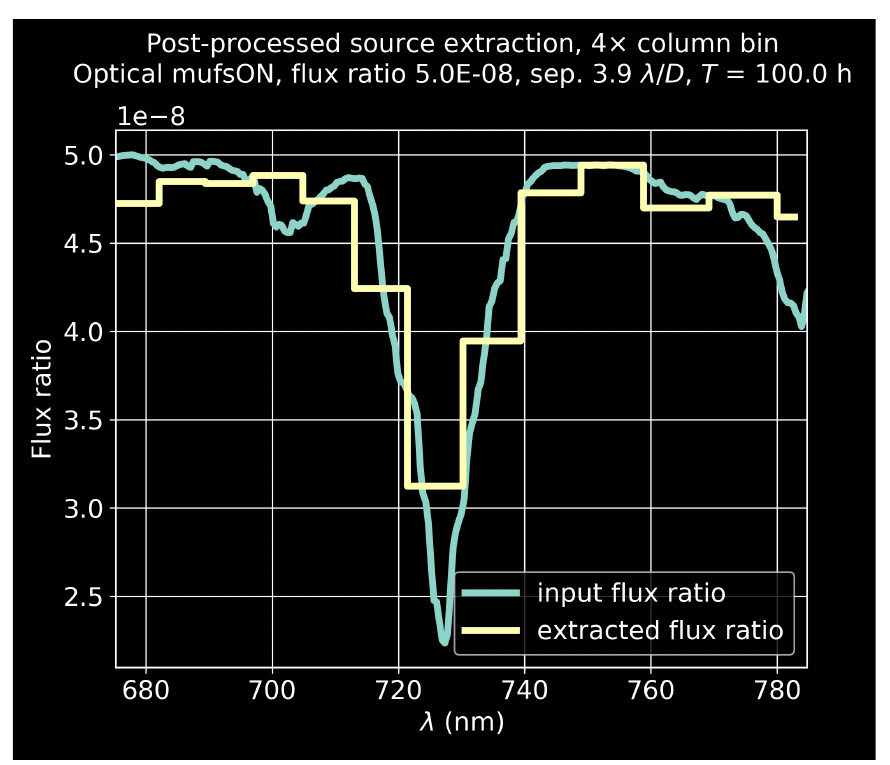}\label{fig:spectrasim}}
    \caption{Simulating the spectrum extraction of a planet around 47 UMa with the slit spectrograph mode of CGI (Band 3). In this illustration we represented the planet source with a Jupiter-like spectral profile, scaled up in intensity to a flux ratio of $5\times10^{-8}$.}
    \label{fig:spectrometer}
\end{figure}

\begin{wrapfigure}{r}{0.35\textwidth}
\centering
\includegraphics[width=.6\linewidth]{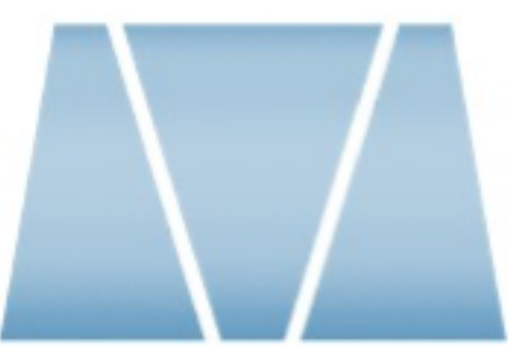}
\caption{Diagram of the Amici prism used for spectral dispersion.}
\label{fig:Amici}
\end{wrapfigure}

While the original plan for CGI called for an integral field spectrograph (IFS) that would provide a full 3D data cube,  cost and schedule constraints forced a descoping of the IFS. CGI is instead equipped with a slit spectrograph, as shown in Figure~\ref{fig:spectrometer}.  The spectrograph will be used in Band 3 with the ``bowtie'' shaped pupil coronagraph. After detection of a planet in Band 1 HLC imaging (and photometric measurements taken), CGI will be reconfigured for Band 3 spectroscopy; the slit will be aligned with the planet as shown in Figure~\ref{fig:slitsim}, which is a simulated spectral image of a $5\times10^{-8}$ flux ratio Jupiter-like planet around 47 UMa.  The planet image is then dispersed across a 15\% band using an Amici prism with an R of 50 at the center of the band, as shown in Figure~\ref{fig:dispersion} (a schematic of the Amici prism is shown in Figure~\ref{fig:Amici}).  Figure~\ref{fig:spectrasim} shows a simulated, post-processed spectrum, clearly indicating the capability to measure a deep methane band.  The spectrometer is equipped with a second prism for spectroscopy in the untested Band 2 as well, should observation time be available.  More detail on spectrometer progress can be found in the paper by Groff.\cite{Groff20}

\begin{figure}[h]
    \centering
    \subfigure[Assembled Prism Module]{\includegraphics[scale=0.6]{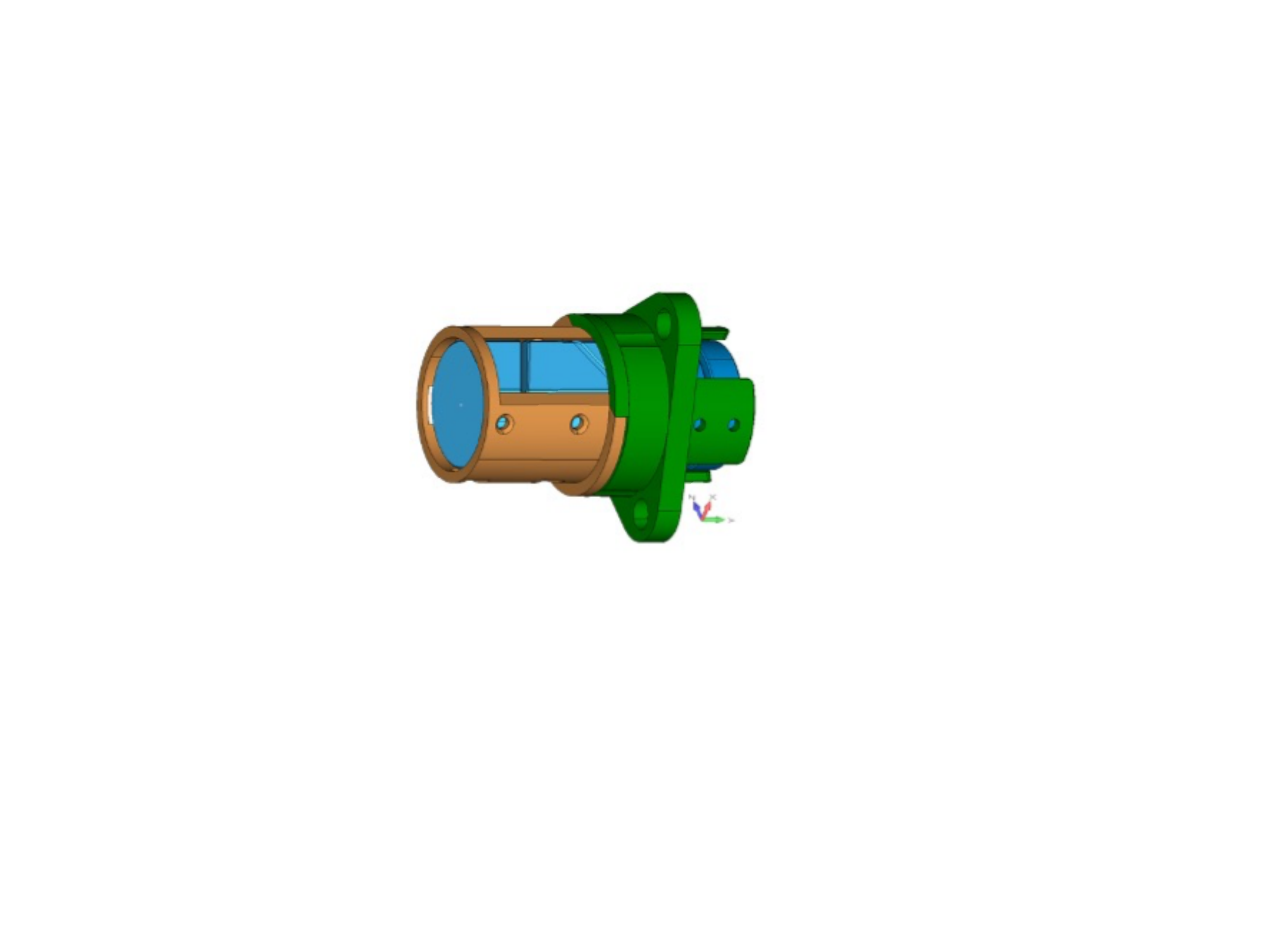}\label{fig:prismmodule}}
    \subfigure[JAXA Polarization Module]{\includegraphics[scale=0.6]{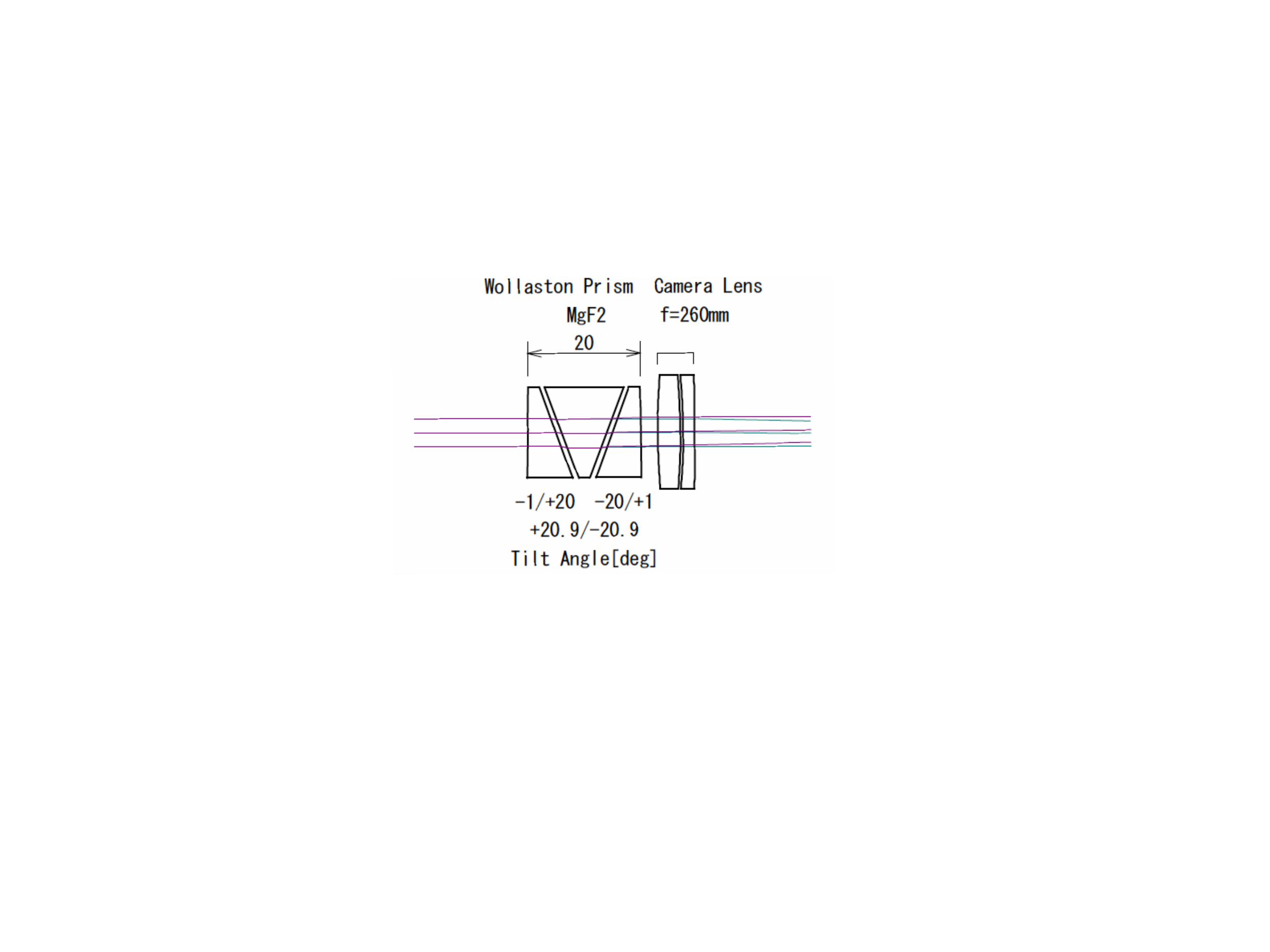}\label{fig:wollaston}}
    \subfigure[Two Pairs of Orthogonally Polarized Images]{\includegraphics[scale=0.6]{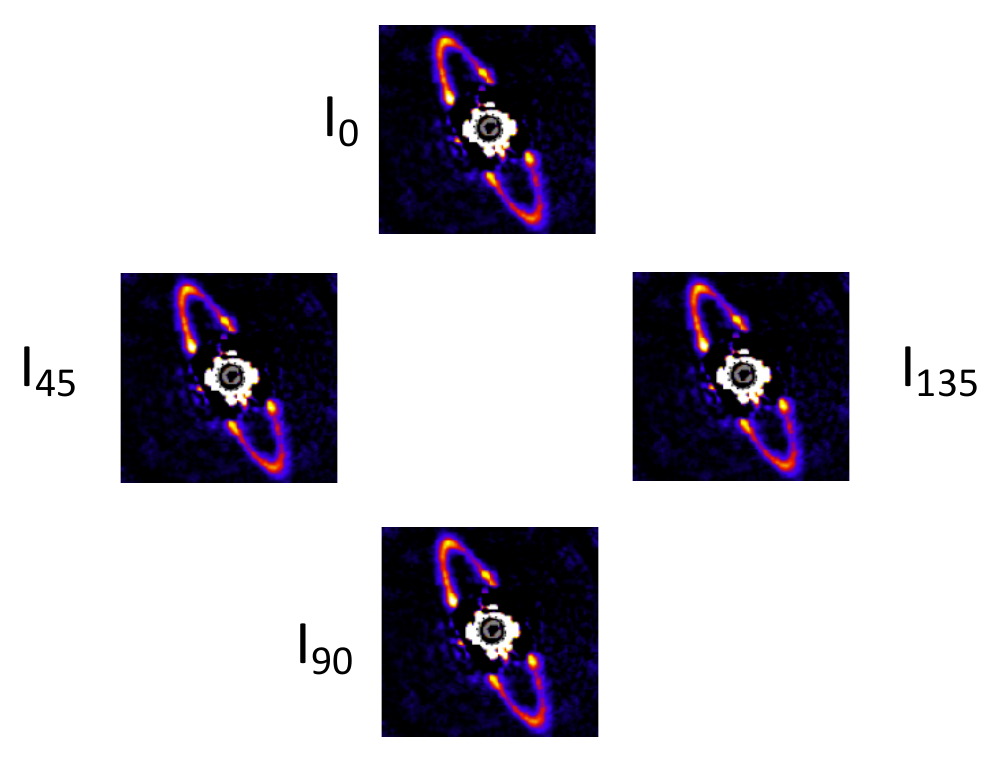}\label{fig:DiskPol}}
    \caption{The Wollaston Prism assembly and sample performance. Only one Wollaston can be used at a time; each produces one pair of images ($0^\circ/90^\circ$ or $45^\circ/135^\circ$).}
    \label{fig:polarization}
\end{figure}

In addition to taking spectra, CGI will measure the linear polarization fraction (LPF), and hence the phase scattering function, of extended objects with an uncertainty on the LPF of better than 3\% (rmse) per spatial resolution element. It will conduct these measurements in CGI bands 1 and 4, which are again centered at 575 and 825~nm, respectively. Polarimetric measurements will be conducted using two Wollaston prisms contributed by JAXA, the Japanese space agency, shown in Figure~\ref{fig:wollaston}.  They are mounted in the polarization module shown in Figure~\ref{fig:prismmodule}.  Each Wollaston prism produces two orthogonally polarized images at a time ($0^\circ/90^\circ$ or $45^\circ/135^\circ$), for a total of four images, as shown in Figure~\ref{fig:DiskPol}. The two Wollastons cannot be used simultaneously, so two sets of observations are needed to collect all four images.

In order to make such measurements, all instrumental polarization effects affecting the overall optical system will be calibrated through observations of (un)polarized standards. It is expected that after this calibration, residual polarization measurement errors will be dominated by detector effects, such as flat-fielding errors. Calibrated polarized images of point-like targets will also be used to validate polarization models that predict different dark hole depths in polarized vs.\ unpolarized light.  These polarization measurements will both provide new and important science, as well as demonstrate for the first time high-contrast polarization capability.

\section{Operations and Image Post-Processing}

\begin{figure}[t]
\centering
\includegraphics[width=.98\linewidth]{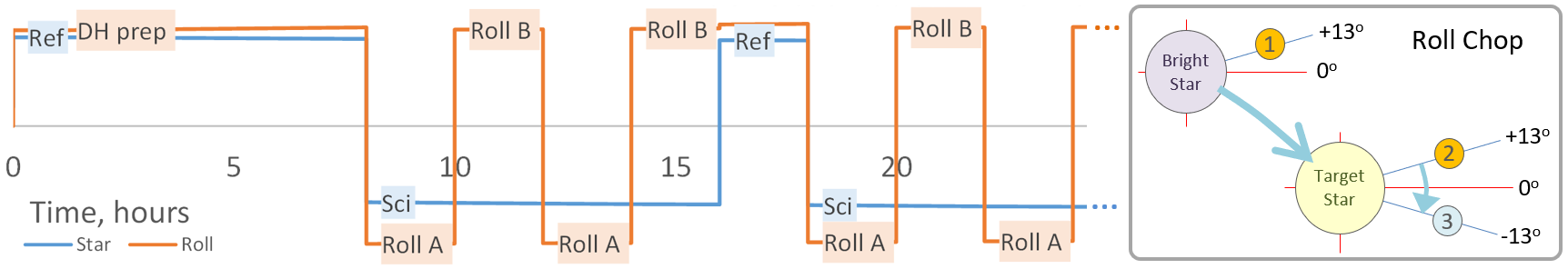}
\vspace{6pt}
\caption{Example observing scenario with $\pm13$ degree roll every 2 hours and slewing to the reference star every 10 hours.}
\label{fig:Observations}
\end{figure}

While achieving high-contrast has been demonstrated by simulation and laboratory experiments, long-term stability of the system still needs to be managed.  Because of potential drifts in the DM and optics, the dark hole needs to be periodically reset.  Additionally, reference images from a bright star are needed to perform PSF subtraction to reveal the planets (known as Reference Differential Imaging, or RDI).  Also, in addition to RDI, Angular Differential Imaging (ADI) is planned to be performed as well.  Hence, the example operational scenario shown Figure~\ref{fig:Observations}.  Every two hours the spacecraft will be rolled by ± 13 degrees and every 10 hours it will be slewed to a reference star to recreate the dark hole and take reference PSFs.  These times are subject to change as more is learned about the stability of the system and as dark hole touch-up cadence is adjusted to match ground station availability in the new HOWFS GITL paradigm.

\begin{figure}
\centering
\includegraphics[width=.5\linewidth]{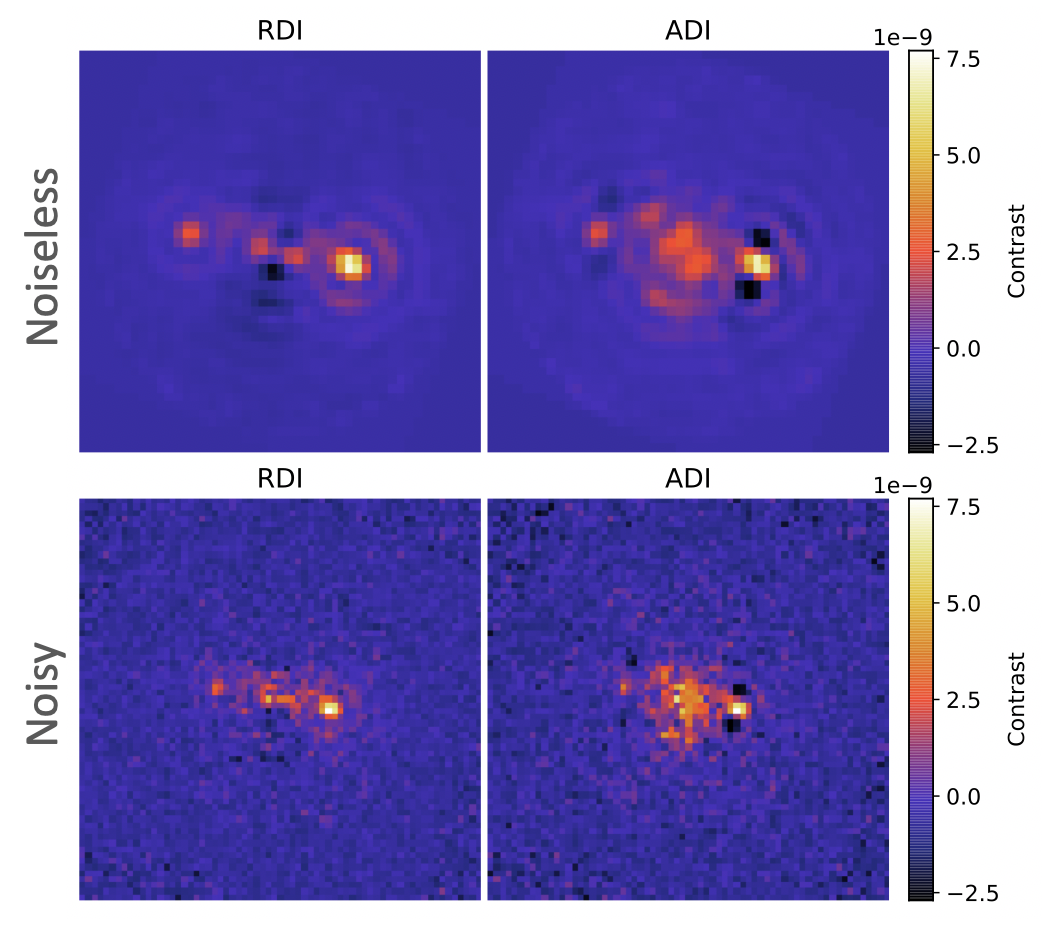}
\vspace{6pt}
\caption{Demonstrating Classical ADI and RDI PSF subtraction on noiseless and noisy simulated data with MUF and with planets (flux ratios of $1\times10^{-8}$ and $3\times10^{-9}$ at $3.5\lambda/D$ and $4.5\lambda/D$ respectively) from Observing Scenario 9 with the Band 1 HLC. }
\label{fig:PSFSubtraction}
\end{figure}

A critical step in image processing is ``post-processing''---the PSF subtraction process to remove the residual speckles in the high-contrast images. Extensive effort has been put into developing and analyzing this approach on simulated datasets.  Simulated images are produced based on the current end-to-end opto-mechanical simulation of the spacecraft and instrument for a realistic target-reference Observing Sequence (OS)\cite{Krist17, Krist18}.  Since before PDR, a sequence of detailed simulations using Structural, Thermal, Optical (STOP) modeling of the observatory, jitter modeling of the Attitude and Translational Control System (ATC), and internal modeling of the CGI coronagraph and wavefront control have been performed, including model uncertainty factors (MUFs), each building on the last by adding more detail.  These end-to-end models produce simulated final images with planets injected including both roll and chopping to the reference star.  These images are used by the team to test post-processing algorithms for detecting planets and to determine the ultimate performance of CGI.\cite{Douglas20}  Images are also made publicly available to allow the community to test alternative approaches.\footnote{CGI simulations available at: \url{https://roman.ipac.caltech.edu/sims/Coronagraph\_public\_images.html}}  Shown in Figure~\ref{fig:PSFSubtraction} are the results of the latest round of tests comparing RDI and ADI approaches on both noiseless and noisy images from the OS9 release.

\section{Science}

\begin{figure}
    \centering
    \includegraphics[scale=0.75]{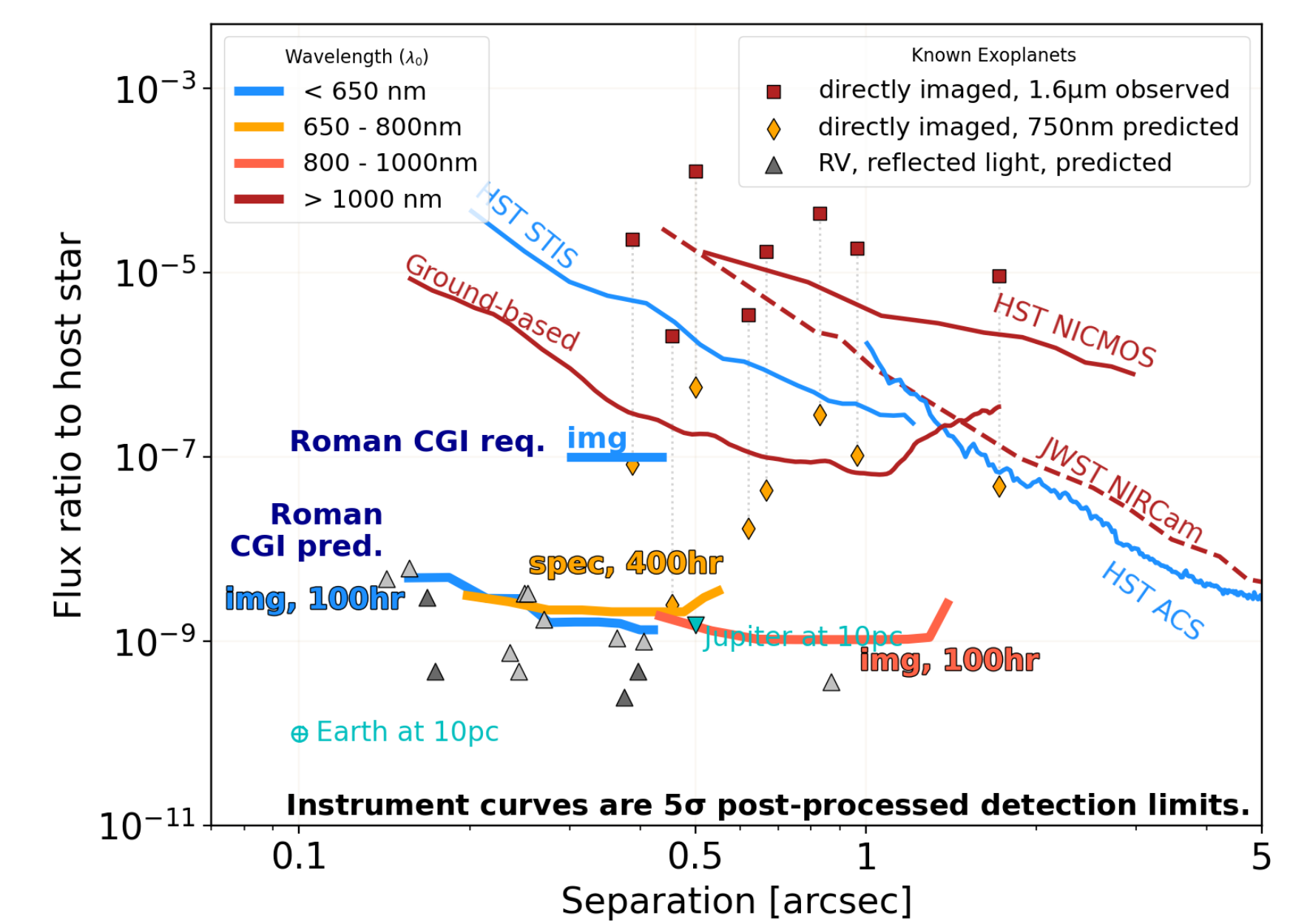}
    \vspace{6pt}
    \caption{CGI's capability for detecting planets at $5\sigma$ as a function of planet-to-star flux ratio and separation from the parent star. CGI provides a 100 to 1000 times improvement over existing and planned facilities.}
    \label{fig:capability}
\end{figure}

The final result of both modeling and experiments is rolled up into the chart in Figure~\ref{fig:capability}.\footnote{plot data is publicly available \url{https://github.com/nasavbailey/DI-flux-ratio-plot}}  This chart shows the predicted capability of CGI in its three bands compared to current ground and space coronagraphs.  Also shown is the current CGI threshold requirement of $10^{-7}$ contrast, the value indicating a successful technology demonstration.  CGI will perform anywhere from 100 to 1000 times better than any current facility.  Current modeling predictions show that CGI performs significantly better than requirements in all bands.  If CGI successfully meets these predictions, significant exoplanet and disk science will be possible, including, as shown, visible light imaging of known young, self-luminous, hot gas-giants and the first reflected light images of Jupiter analogs, including current nearby RV targets.

\begin{figure}
    \centering
    \subfigure[Self-Luminous Jupiters]{\includegraphics[scale=1.2]{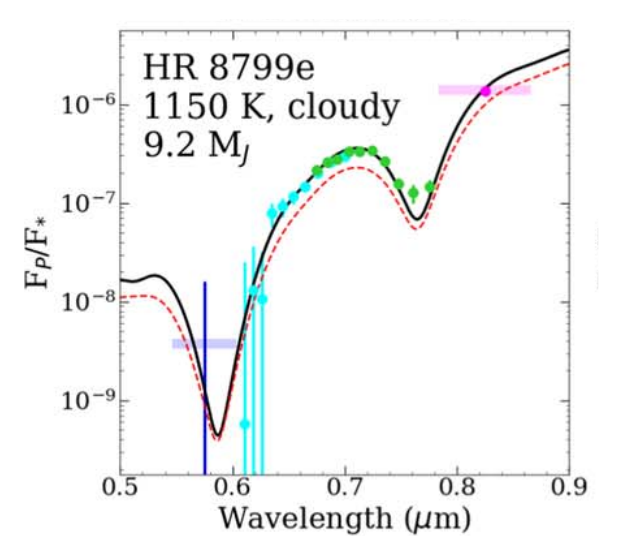}\label{fig:8799e}}
    \subfigure[Reflected Light Jupiters]{\includegraphics[scale=0.35]{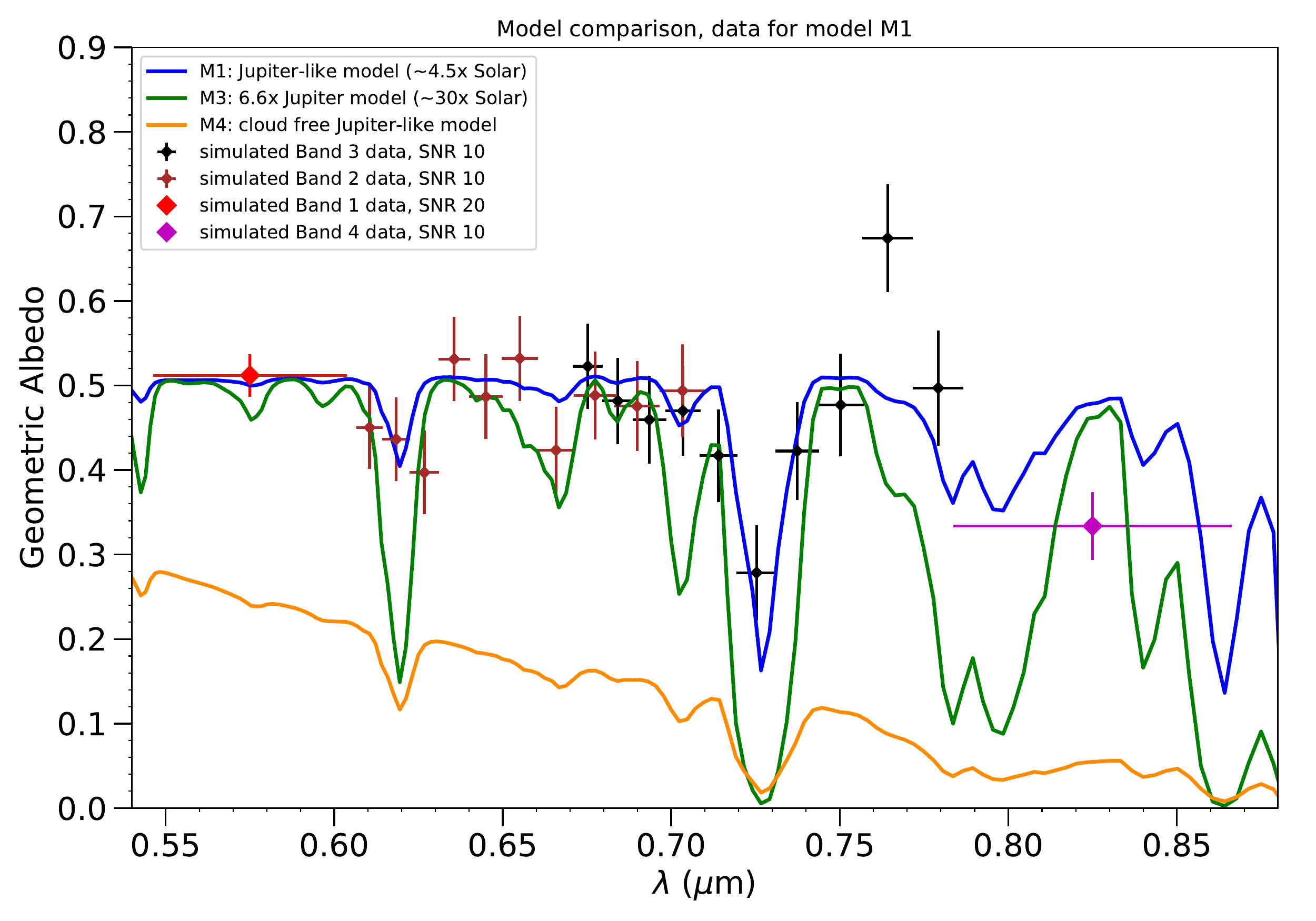}\label{fig:jupiter}}
    \subfigure[Circumstellar Disks]{\includegraphics[scale=0.4]{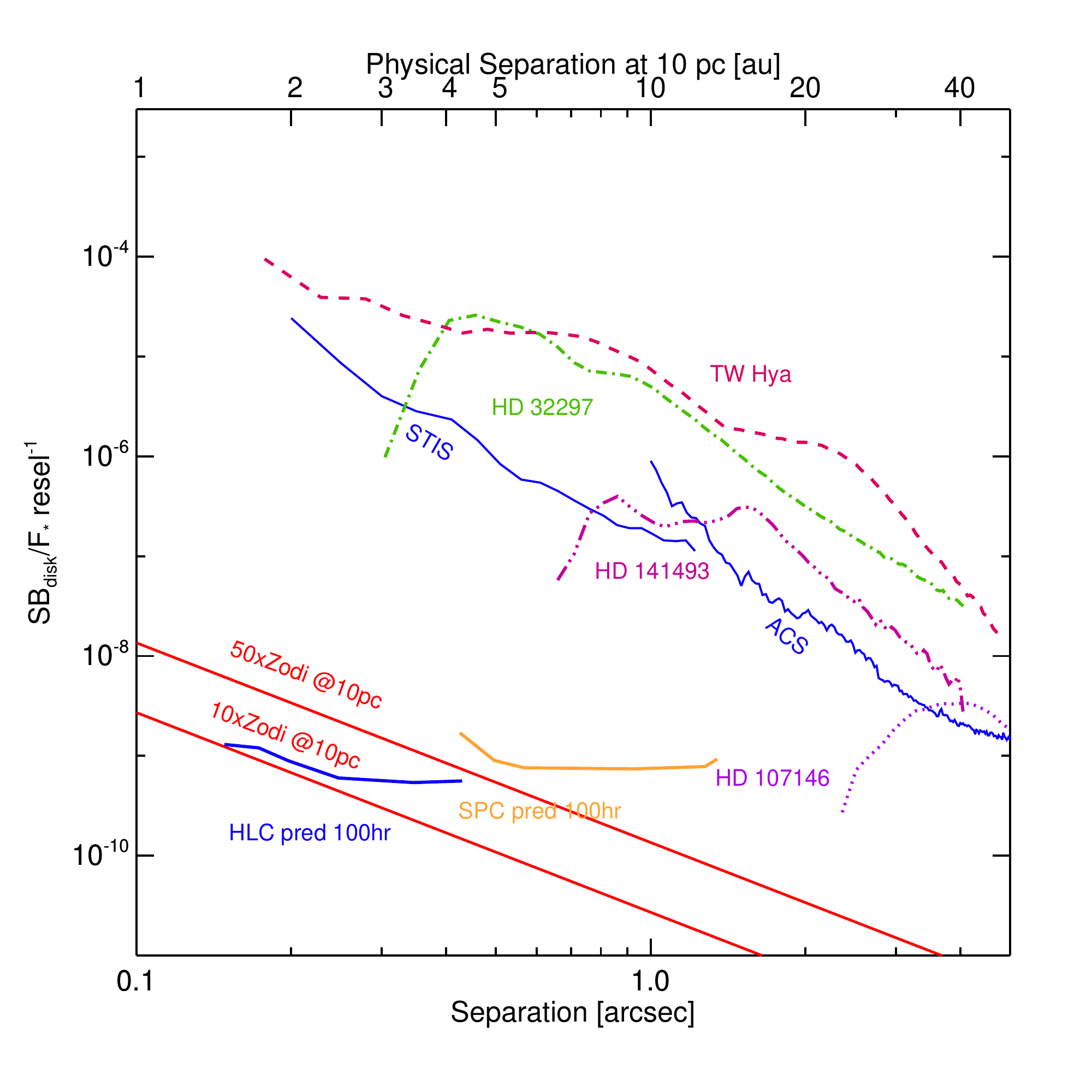}\label{fig:disks}}
    \caption{Three sample science cases for CGI after the technology demonstration phase. (a) Spectra of young, hot, self-luminous Jupiters to determine atmospheric properties. \cite{lacy20} (b) Atmospheric properties of mature Jupiter-analogues in reflected light to determine their albedo. (figure courtesy of Roxana Lupu - NASA Ames)  (c) Photometric characterization of all three types of circumstellar disks: proplanetary (young; TW Hydra), debris (mature; dashed curves), and exozodi (solid red lines). Roman/CGI's HLC (shown lower left) will provide higher precision observations of these objects than those currently capable with Hubble's STIS and ACS (shown upper right). (figure courtesy of John Debes - STScI.)}
    \label{fig:Science}
\end{figure}

\begin{figure}
    \centering
    \includegraphics[scale=0.8]{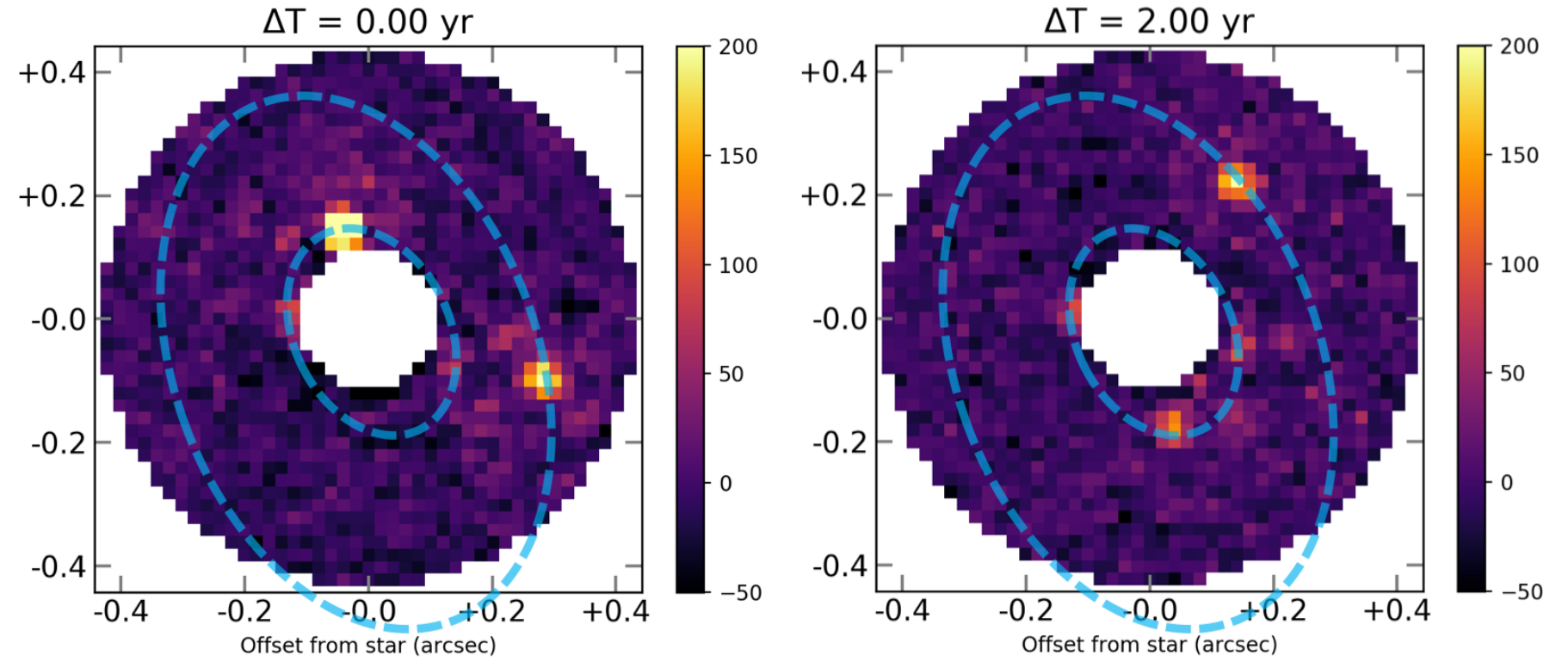}
    \caption{Simulated, post-processed CGI HLC images of a multiplanetary system observed at two epochs separated by two years, displayed in units of integrated photoelectrons. The blue, dashed ellipses illustrate the projected orbits of the two planets detected as point sources. These simulations were produced using ``OS6'' integrated models that include the effects of time-varying spacecraft pointing jitter, low-order aberrations, and speckle noise. Similar multi-epoch data sets were simulated for a community data challenge organized by one of the science teams, which probed strategies for retrieving planet orbits and masses\cite{Girard20}.}
    \label{fig:Orbit}
\end{figure}

While CGI is designated as a technology demonstrator, the team is confident that its performance will exceed requirements, allowing us to perform new and exciting exoplanet science both during and after the initial technology demonstration phase.  As shown in Figures~\ref{fig:Science} and \ref{fig:Orbit}, there is a robust set of science possible by observing young, self-luminous, gas giants\cite{lacy20} as well as direct imaging of mature Jupiter analogs in visible, reflected light.\footnote{Predicted flux ratios for RV planets are publicly available via the Imaging Mission Database\cite{Wenger2000, Batalha2018, Turnbull2015}: \url{https://plandb.sioslab.com/}}  Multiple observations will also allow orbit determination of imaged planets.  CGI will also be capable of direct imaging and polarization measurements of circumstellar disks, particularly debris disks, as well as potentially characterizing protoplanetary disks and taking the first visible-light image of an exozodi disk.  

Current ground-based observations are exclusively limited to observing young, hot super-Jupiters in the infra-red relatively far from their host star.  Roman/CGI will take visible-light images and spectra of these planets, enabling the distinction between clear and cloudy atmospheres and further constraining their properties and formation mechanisms\cite{lacy20}.  Shown in Figure~\ref{fig:8799e} is the planet to star contrast as a function of wavelength for the young, self-luminous giant planet HR~8799e assuming an equilibrium temperature of 1150~K, a cloudy atmosphere, and a mass of 9.2~Jupiter masses. The CGI passbands are also shown. While infrared wavelengths already favor a cloudy rather than a clear atmosphere for HR~8799e, adding optical coverage with CGI should improve our understanding of the cloud properties beyond just stating that they are there (e.g. rule out/favor some chemical compositions, provide more information on particle sizes). Optical coverage with CGI should tighten constraints on other planet properties as well (e.g., planetary metallicity).

Roman/CGI will also take the first visible-light images and spectra of cool, true Jupiter analogs (old and cold Jupiter-mass planets orbiting several AU from their stars), thus coarsely probing the metallicity and cloud properties of Jupiter analogues, as illustrated in Figure~\ref{fig:jupiter}.  Shown as solid lines are 3 models: M1---a Jupiter-like planet model with a $\sim$4.5$\times$ Solar metallicity; M3---a 6.6$\times$ Jupiter model ($\sim$30$\times$ Solar Metallicity); and M4---a cloud-free Jupiter-like model. Simulated CGI observations for model M1 are shown as datapoints for Bands 1 (photometry; red) and 3 (spectroscopy; black).  CGI could loosely constrain metallicity (solar vs.\ enriched) for the cloudy planets, providing information important to  formation modeling and as precursors to future missions both by demonstrating high-contrast spectroscopy and highlighting potential targets.

CGI can probe the formation and evolution of extrasolar systems by observing all three types of circumstellar disks at lower density and in regions closer to the star than currently possible from the ground, as shown in Figure~\ref{fig:disks}.   Imaging of debris disks provides information on the remains of planet formation as well as colliding or evaporating asteroids and comets. Polarimetric observations complement integrated light imaging to constrain grain properties. Roman/CGI could also provide the first close-in, visible-light images of exozodiacal dust clouds produced by collisions of asteroids and comets.  These clouds can potentially obscure small Earth-like planets; CGI images will provide crucial information for future mission planning. If CGI WFS/C is able to perform well on fainter target stars, it could observe protoplanetary and transition disk systems, providing information on newly-forming planetary systems.  Figure~\ref{fig:disks} shows predicted CGI disk imaging performance in the context of HST detection limits and surface brightness profiles of previously-resolved disks and exozodi predictions. Surface brightness units are flux ratio per resolution element.   

CGI will also measure the orbital solution of planets that have been previously detected via the radial velocity method, allowing the direct measurement of a planet’s orbital inclination, providing a crucial constraint on m*sin(i), and allowing us to directly measure the planet’s mass. (See Figure~\ref{fig:Orbit}.)

\section{Timeline}

CGI is on track to meet its schedule targets.  Great progress has been made since entering Phase C last February with almost all engineering units procured and many flight units in manufacture.  The team anticipates a successful CDR this coming April and delivery of the completed instrument to the Goddard Space Flight Center for integration and test in 2023.  After launch in late 2025, CGI will enter a commissioning phase during the first 90 days, followed by its 3 month technology demonstration phase spread across the first 1.5 years of the mission.  Upon completion of the technology demonstration phase, if deemed successful, and if additional funding is made available, CGI could potentially be available for science- and technology-driven observations using somewhere between 10\% and 25\% of the remaining Roman mission time.  

A competitively selected Community Participation Program is planned to be  an integral part of the technology demonstration and could lead additional CGI observations beyond the initial tech demo if the instrument performance warrants it. The Community Participation Program will be the mechanism through which members of the community engage in both the technology demonstration phase and potential science phase.  Details of the program are still being worked out and a call for applications is expected sometime in 2021.   

\section{Summary}

In Summary, the CGI instrument on the Roman space telescope is on schedule for CDR in April 2020 and launch in 2025 and is exceeding its performance requirements.  Upon launch and commencement of operation, it will be the first actively controlled, high-contrast coronagraph in space.  Its combination of multiple technologies (including three different coronagraphs) and modes of operation while raising the TRL (technology-readiness level)  of critical hardware, such as deformable mirrors, low-noise, photon-counting detectors, coronagraph masks, and other elements will make CGI a critical technology demonstration for future missions, paving the way for direct imaging of Earth-like planets in the habitable zone of nearby stars.  If successful, CGI will also provide breakthrough science, including spectroscopy of known RV and self-luminous planets, imaging and photometry of reflected light planets, and characterization of zodiacal dust and debris disks.  

\section{Acknowledgements}
Part of the work described here was carried out at the Jet Propulsion Laboratory, California Institute of Technology, under contract with the National Aeronautics and Space Administration. Copyright 2020. All rights reserved.  Work also supported by a grant from NASA Goddard Space Flight Center, Award \#NNG16PJ29C. This research has made use of the Imaging Mission Database, which is operated by the Space Imaging and Optical Systems Lab at Cornell University. The database includes content from the NASA Exoplanet Archive, which is operated by the California Institute of Technology, under contract with the National Aeronautics and Space Administration under the Exoplanet Exploration Program, and from the SIMBAD database, operated at CDS, Strasbourg, France.

\bibliography{KasdinSPIEbib} 
\bibliographystyle{spiebib} 

\end{document}